\documentclass[preprint,showpacs,preprintnumbers,amsmath,amssymb]{revtex4}

\usepackage{graphicx}
\usepackage{dcolumn}
\usepackage{bm}

\begin{document}

\preprint{APS/123-QED}

\title{Deformation of hypernuclei studied with antisymmetirzed molecular dynamics}
\author{M. Isaka}
\affiliation{
Department of Cosmosciences, Graduate School of Science, Hokkaido University,
Sapporo 001-0021, Japan}
\author{M. Kimura}
\affiliation{
Creative Research Institution (CRIS), Hokkaido University, Sapporo 060-0810, Japan}
\author{A. Dote}
\affiliation{
Institute of Particle and Nuclear Studies, KEK, Tsukuba, Ibaraki 305-0801, Japan}
\author{A. Ohnishi}
\affiliation{
Yukawa Institute for Theoretical Physics, Kyoto University, Kyoto 606-8502,
Japan}

\date{\today}

\begin{abstract}
 An extended version of the antisymmetrized molecular dynamics to study structure of
 $p$-$sd$ shell hypernuclei is developed. By using an effective $\Lambda N$
 interaction, we  investigate energy curves of $^9_\Lambda$Be,  $^{13}_\Lambda$C and  
 $^{20,21}_\Lambda$Ne as function of nuclear quadrupole deformation. Change of
 nuclear deformation caused by $\Lambda$ particle is  discussed. It is found  that
 the $\Lambda$ in $p$-wave enhances nuclear  deformation,  while that in $s$-wave
 reduces it. This effect is most prominent in  $^{13}_\Lambda$C. The possibility of
 the parity inversion in $^{20}_\Lambda$Ne is  also examined.
\end{abstract}

\pacs{Valid PACS appear here}
\maketitle

\section{Introduction}
One of the unique and interesting aspects of hypernuclei is the structure change
caused by the hyperon as an impurity. Many theoretical works have suggested such
phenomena caused by  $\Lambda$ particle as the change of deformation
\cite{Motoba83,Motoba85,Zhou07,Win08,Shulze10}, the 
shrinkage of the inter-cluster distance \cite{Motoba83,Motoba85} and the
super-symmetric (genuine) hypernuclear state
\cite{Dalitz76,Motoba85,Sakuda87,Straub88,Yamada88}. Owing to the 
experimental developments some of them have been observed in light $p$-shell
hypernuclei. As examples, we can refer to the reduction of B$(E2)$ in $^7_\Lambda$Li
\cite{Selove88} and the identification of the super-symmetric (genuine) hypernuclear
state in $^9_\Lambda$Be \cite{Bertini81-1, Has98, Hashimoto06}. 

Today, we can expect that a new experimental facility of Japan Proton Accelerator 
Research Complex (J-PARC) will reveal the spectral information of $p$-$sd$ shell and
neutron-rich hypernuclei. Since these normal nuclei have a variety of structure
such as coexistence of shell and cluster structure \cite{Enyo98,Kimura04,Kimura06}
and novel exotic clustering \cite{Oertzen97,Enyo95,Itagaki00,Desc02,Enyo02},
there must be many interesting phenomena peculiar to hypernuclei to be
found. Indeed, several pioneering works predicted exotic structure in $sd$-shell
hypernuclei such as the parity inversion in $^{20}_\Lambda$Ne \cite{Sakuda87} and
various rotational bands in  $^{21}_\Lambda$Ne \cite{Yamada84}.
 These works are based on rather limited knowledge on the 
$\Lambda$N interaction. Since our knowledge of $\Lambda$N interaction has
been greatly increased by the recent theoretical and experimental efforts 
\cite{Reuver94,Rijken06,Fujiwara07,Millener08,Hiyama09,Hashimoto06},
we are now able to perform more quantitative and systematic study of the structure 
change in $\Lambda$ hypernuclei. It will reveal the dynamics and many interesting
aspects of baryon many-body problem.

To perform systematic and quantitative study of $sd$-shell and neutron-rich
$\Lambda$ hypernuclei, we develop an extended version of the antisymmetrized
molecular dynamics (AMD) \cite{Enyo95,Dote97,Sugawa01,Kimura01,Enyo03}, which we shall call HyperAMD. AMD has been successful to describe various
exotic structure of neutron-rich nuclei and highly excited states of stable
nuclei. Therefore, HyperAMD is suitable to describe the structure change and exotic
structure of hypernuclei.  

In this study, we introduce HyperAMD and focus on the change of nuclear quadrupole
deformation caused by a 
$\Lambda$ particle. By applying HyperAMD to $^{9}_\Lambda$Be, $^{13}_\Lambda$C,
$^{20}_\Lambda$Ne and $^{21}_\Lambda$Ne with YNG-ND $\Lambda$N interaction
\cite{Yamamoto94}, it is found that $\Lambda$ particle changes nuclear quadrupole
deformation.  While the $\Lambda$ particle in $s$-wave reduces quadrupole
deformation as expected, that in $p$-orbital increases it. Among the calculated
hypernuclei, $^{13}_\Lambda$C has shown the most drastic change of the nuclear
deformation. It is also found that the binding energy of $\Lambda$ particle depends
on the structure of the core nucleus. Namely, the $\Lambda$ in $s$-wave coupled to
the deformed core nucleus has smaller binding than that coupled to spherical core. On
the contrary, the $\Lambda$ in $p$-wave coupled to the deformed core has larger
binding than that coupled to the spherical core. This contradicts to the preceding
study \cite{Sakuda87} in which the parity inversion of $^{20}_\Lambda$Ne is
predicted.

This paper is organized as follows. In the next section, we explain the
theoretical framework of HyperAMD. In the section III, the change of energy curves
as function of quadrupole deformation are presented. The trend of the change and its
origin are discussed. The final section summarizes this work.

\section{Theoretical Framework}
In this section, we introduce the theoretical framework of
HyperAMD. Compared to the coupled channel AMD which describes the multi-strangeness
system \cite{Matsumiya10}, it has better description of the hyperon single particle wave
function, but it does not treat multi-strangeness and is limited to single
$\Lambda$ hypernuclei. 

\subsection{Wave function}
A single $\Lambda$ hypernucleus consists of $A$ nucleons and a $\Lambda$ particle
is described by the wave function that is the eigenstate of the  parity, 
\begin{eqnarray}
 \Psi^{\pm} = \hat{P}^{\pm}\Psi_{\rm int},
\end{eqnarray}
where $\hat{P}^{\pm}$ is the parity projector, and the intrinsic wave function
$\Psi_{\rm int}$  is given by the direct product of the $\Lambda $ single particle
wave function $\varphi_\Lambda$ and the wave function of $A$ nucleons $\Psi_N$,  
\begin{eqnarray}
\Psi_{\rm int} &=& \varphi_\Lambda\otimes \Psi_{N}.
\end{eqnarray}
The nuclear part is described by a Slater determinant of nucleon single particle
wave packets,
\begin{eqnarray}
\Psi_N &=& \frac{1}{\sqrt{A!}}\det \left\{ \psi_{i}(\bm r_j) \right\},\\
\psi_{i}(\bm r_j) &=& \phi_{i}(\bm r_j)\cdot \chi_{i}\cdot \eta_{i},\\
\phi_{i}(\bm r) &=& \prod_{\sigma=x,y,z} \biggl(\frac{2\nu_\sigma}{\pi}\biggr)^{1/4}
 \exp \biggl\{-\nu_\sigma \bigl(r - Z_{i} \bigr)_\sigma^2 \biggr\},\\ 
\chi_{i} &=& \alpha_i \chi_\uparrow + \beta_i \chi_\downarrow,\\
\eta_{i} &=& {\rm proton}\ {\rm or}\ {\rm neutron},
\end{eqnarray}
where $\psi_{i}$ is the {\it i}-th nucleon single-particle wave packet consists of
spatial $\phi_{i}$, spin $\chi_{i}$ and isospin $\eta _{i}$ parts. The spatial part 
$\phi_{i}$ is represented by a deformed Gaussian. Its centroid ${\bm Z}_i$ is a
complex valued three-dimensional vector. The width parameters $\nu_{\sigma}$ are
real numbers and take independent value for each direction, but are common to all
nucleons. The spin part is parameterized by the complex parameters $\alpha_i$ and
$\beta_i$, and the isospin part is fixed to proton or neutron. $\bm{Z}_i$,
$\nu_\sigma$, $\alpha_i$ and $\beta_i$ are the variational parameters of the nuclear
part.  

To describe various wave functions of the $\Lambda$ particle, the $\Lambda$ single
particle wave function is represented by the superposition of  Gaussian wave
packets, 
\begin{eqnarray}
\varphi_\Lambda(\bm r) &=& \sum_{m=1}^M c_m \varphi_m(\bm r),\quad 
\varphi_m(\bm r) = \phi_m(\bm r)\cdot \chi_m,\\
\phi_m(\bm r) &=& \prod_{\sigma=x,y,z} \biggl(\frac{2\nu_\sigma\rho}{\pi}\biggr)^{1/4}
 \exp \biggl\{-\nu_\sigma \rho\bigl(r - z_m \bigr)_\sigma^2 \biggr\},\\ 
 \chi_m &=& a_m \chi_\uparrow + b_m \chi_\downarrow,\\
 \rho &\equiv& \frac{m_\Lambda}{m_N}.
\end{eqnarray}
Again, each wave packet is parametrized by the centroid of Gaussian $\bm{z}_m$, the
spin direction $a_m$ and $b_m$. $\bm{z}_m$, $a_m$, $b_m$ and $c_m$ are   
the variational parameters of the $\Lambda$ part. The width parameter
$\nu_\sigma$ are equal to those of nuclear part. The number of the superposition
$M$ is taken to be large enough to have the energy convergence of the variational
calculation. 

\subsection{Hamiltonian}
The Hamiltonian used in this study is given as 
\begin{equation}
\hat{H} = \hat{T}_{N} + \hat{V}_{NN}  + \hat{V}_{Coul} 
 + \hat{T}_{\Lambda} + \hat{V}_{\Lambda N} - \hat{T}_g.
\end{equation}
Here, $\hat{T}_{N}$, $\hat{T}_{\Lambda}$ and $\hat{T}_g$ are the kinetic energies of
nucleons, $\Lambda$ particle and the center-of-mass motion. Since we have
superposed Gaussian wave packets to describe the $\Lambda$ single particle wave
function, it is rather time consuming to remove the spurious motion of the
center-of-mass exactly. To reduce the spurious motion, we keep the center-of-mass of
wave packets at the origin of the coordinate,  
\begin{eqnarray}
\sum_{i=1}^{A} {\bm Z}_i + \sum_{m=1}^M\sqrt{\rho} {\bm z}_m= 0.
\end{eqnarray}
We expect that the spurious energy is not large in $sd$-shell hypernuclei, since the
number of nucleons is much larger than the hyperon. A similar method is also  applied
in the other AMD studies \cite{Dote97}.   

Our model wave function is designed to describe the low-momentum phenomena as in the
case of the conventional shell model and we shall use the low-momentum effective
interaction. We have used the Gogny D1S interaction \cite{Gogny80} as an effective
nucleon-nucleon interaction $\hat{V}_{NN}$, that has been successfully applied to
the stable and unstable normal nuclei. As an effective $\Lambda$N
interaction, we have used the central part of the YNG-ND interaction
\cite{Yamamoto94}. The YNG-ND interaction depends on the nuclear Fermi momentum $
k_{F} $ through the  density-dependence of the G-matrix in nuclear medium. In this
work, we apply respectively $k_F$=1.14 and 1.17 fm$^{-1}$ for $^{9}_\Lambda$Be and 
$^{13}_\Lambda$C, that are so determined to approximately reproduce the binding energy of 
$\Lambda$ in $s$-wave. For $^{20}_{\Lambda}$Ne and  $^{21}_{\Lambda}$Ne, we apply
the same value as $^{13}_{\Lambda}$C, since there is no experimental data.  The
Coulomb interaction is approximated by the sum of seven Gaussians.

\subsection{Frictional cooling method with constraints}
Using the frictional cooling method, the variational parameters of the model wave
function are so determined that the total energy is minimized under the
constraints. We have imposed two constraints on the variational calculation. The
first is on the nuclear quadrupole deformation parameter $\beta$ that is
achieved by adding the parabolic potential, 
\begin{eqnarray}
 \langle\hat{V}_\beta\rangle = v_\beta (\beta - \beta_0)^2,
\end{eqnarray}
to the total energy. Here $\beta$ denotes the quadrupole deformation
of the nuclear wave function $\Psi_N$ \cite{Dote97}. The  deformation of nuclear
part  becomes equal to $\beta_0$ after the variation. It is
noted that  there are no constraint on the nuclear quadrupole deformation $\gamma$
and the deformation of the $\Lambda$ single particle wave function. They have the
optimum value after the variational calculation for each given value of $\beta_0$. 

Another constraint is on the $\Lambda$ single particle wave function, 
\begin{eqnarray}
\hat{V}_{s} &=& \Lambda |\varphi_s\rangle\langle \varphi_s|,\\
 \langle \bm{r}|\varphi_s\rangle &=& \exp\{-\rho\bar{\nu} r^2\},\\
 \bar{\nu} &=& \sqrt[3]{\nu_x\nu_y\nu_z}
\end{eqnarray}
By using sufficiently large value for $\Lambda$, this potential forbids the
$\Lambda$ in $s$-wave. Therefore, by switching off and on this potential,
we respectively obtain the hypernuclear state in which a $\Lambda$ particle 
dominantly occupies in $s$- and $p$-waves.

The total energy plus constraint potentials, 
\begin{eqnarray}
 E^\prime = \frac{\langle \Psi^\pm|\hat{H}|\Psi^\pm\rangle}{\langle
  \Psi^\pm|\Psi^\pm\rangle} + \langle \hat{V}_s\rangle + \langle
  \hat{V}_\beta\rangle 
\end{eqnarray}
is minimized using the frictional cooling method. The imaginary time development
equations of the variational parameters are given as,
\begin{eqnarray}
 \frac{dX}{dt} &=& \frac{\mu}{\hbar}\frac{\partial E^\prime}{\partial X^*},\\
 X &=& \bm{Z}_i, \bm{z}_m, \alpha_i, \beta_i, a_m, b_m,\nu_\sigma,\nu^\Lambda_i,
\end{eqnarray}
where $\mu$ is arbitrary negative real number. It is easy to proof that $E^\prime$
decreases as time develops, and after sufficient time steps we obtain the energy
minimum under the constraint. By this method, we obtain the hypernuclear wave
function for given $\beta_0$, the total parity and the $\Lambda$ single particle
orbital. In the present work, $\Lambda$ dominantly occupies $s$- or $p$-wave and we shall
use the notation $\Lambda_s$ and $\Lambda_p$ for them. Combined with the parity
projection, we obtain four different configurations 
in which $\Lambda_s$ and $\Lambda_p$ couple to the positive- and negative-parity
states of the core. They are denoted as $\Psi_N^+\otimes \Lambda_s$,
$\Psi_N^-\otimes \Lambda_s$, $\Psi_N^+\otimes \Lambda_p$ and 
$\Psi_N^-\otimes \Lambda_p$ in the following.

\section{Results and Discussions}

\subsection{general trend of the energy curves}

\begin{figure*}[th]
 \includegraphics[width=0.8\hsize]{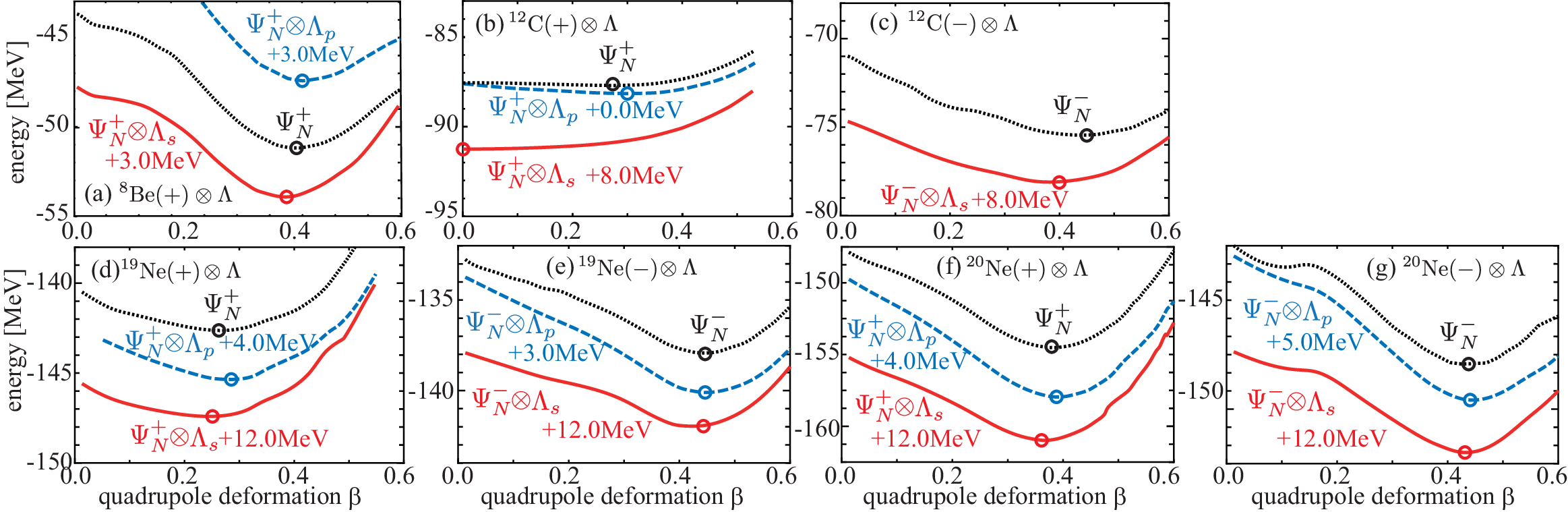}
 \caption{(color online)
 Energy curves as function of nuclear quadrupole deformation $\beta$ for
 (a) $^{9}_\Lambda$Be, (b)(c) $^{13}_\Lambda$C, (d)(e) $^{20}_\Lambda$Ne and
 (f)(g) $^{21}_\Lambda$Ne. (a),(b),(d) and (f) compare the positive-parity states of
 normal nuclei ($\Psi_N^+$) with the hypernuclear states of 
 $\Psi_N^+\otimes\Lambda_s$ and $\Psi_N^+\otimes\Lambda_p$ configurations.
 (c),(e) and (f) compare the  negative-parity states ($\Psi_N^-$) with
 the hypernuclear states of $\Psi_N^-\otimes\Lambda_s$ and
 $\Psi_N^-\otimes\Lambda_p$   configurations.  Open circle shows the energy minimum
 on each  curve. Energies of hypernuclei are shifted as shown in the figure for the
 sake of  the presentation.}\label{fig:energy_curve} 
\end{figure*}

We have performed the variational calculation for  $^{9}_\Lambda$Be,
$^{13}_\Lambda$C, $^{20}_\Lambda$Ne and $^{21}_\Lambda$Ne. To illustrate the change 
of nuclear deformation, Figure \ref{fig:energy_curve} shows energy curves of
hypernuclear states with different configurations and corresponding normal nuclei as
functions of deformation $\beta$. 
Each energy curve has an energy minimum shown by the open circle, and the binding
energies, quadrupole deformation and radius at the minimum are listed in Table 
\ref{tab:min}. The binding energy of $\Lambda$ is defined as the difference of
energy between the ground state of the core nucleus and the hypernuclear states,
\begin{eqnarray}
 B_\Lambda = B(^{A+1}_\Lambda X)  - B(^AX_{g.s.})
\end{eqnarray}

\begin{table}[thb]
\caption{Calculated total and  $\Lambda$ binding energies $B$, $B_\Lambda$ in 
 MeV, quadrupole deformation parameters $\beta$ and $\gamma$, and  the  root mean 
 square radius in fm at the minimum of each energy  curve. Central values of observed
 energies   \cite{Hashimoto06,Jur73,Dav92,Has98} are  also listed in parenthesis. 
 The energies for $\Lambda_p$ states of $^{9}_\Lambda$Be and $^{13}_\Lambda$C
 are estimated from the observed excitation energies given in
 Ref. \cite{Hashimoto06}.}\label{tab:min}  
\begin{ruledtabular}
\begin{tabular}{ccccccc}
& & $B$ & $B_\Lambda$ & $\beta$ & $\gamma$ &$\sqrt{\langle r^2\rangle}$\\
 \colrule
 $^{8}$Be &$\Psi_N^+$ & 51.2 (56.5) & - & 0.68 & 1.9 & 2.50\\
 $^{9}_\Lambda$Be &$\Psi_N^+\otimes\Lambda_s$ &56.9 (63.2)& 5.75 (6.71)  & 0.65 &
		     1.9 & 2.44\\  
 &$\Psi_N^+\otimes\Lambda_p$& 50.4 (56.7) & -0.77 (0.19) & 0.71 & 1.7 & 2.53\\
 \colrule
 $^{12}$C &$\Psi_N^+$ & 87.7 (92.2) & - & 0.27 & 60.0 & 2.42\\
 &$\Psi_N^-$ & 75.5(82.5) & - & 0.45 & 45.4 & 2.56\\
 $^{13}_\Lambda$C &$\Psi_N^+\otimes\Lambda_s$ & 99.3 (103.9) & 11.6 (11.69) & 0.00 &
		     - & 2.32 \\ 
 &$\Psi_N^+\otimes\Lambda_p$& 88.2 (93.8) & 0.46 (1.65) & 0.30 & 55.1 & 2.42 \\
 &$\Psi_N^-\otimes\Lambda_s$ & 86.2 & -1.5 & 0.40 & 42.5
		     &2.49\\ 
 \colrule
 $^{19}$Ne &$\Psi_N^+$ & 142.6(143.7) & - & 0.27 & 0.6 & 2.81\\
 &$\Psi_N^-$ & 137.9(143.5) & - & 0.45 & 0.5 & 2.91\\
 $^{20}_\Lambda$Ne & $\Psi_N^+\otimes\Lambda_s$ & 159.4 & 16.8 & 0.25 & 0.6 & 2.76\\
 &$\Psi_N^+\otimes\Lambda_p$& 148.4 & 5.74 & 0.30 & 0.9 & 2.81\\
 &$\Psi_N^-\otimes\Lambda_s$& 154.0 & 11.4 & 0.45 & 0.5 & 2.87\\
 &$\Psi_N^-\otimes\Lambda_p$& 144.2 & 1.6 & 0.45 & 0.5 & 2.89\\
 \colrule
 $^{20}$Ne &$\Psi_N^+$ & 155.6(160.6) & - & 0.38 & 0.7 & 2.89\\
 &$\Psi_N^-$ & 147.5(155.6)& - & 0.43 & 0.4 & 2.91\\
 $^{21}_\Lambda$Ne &$\Psi_N^+\otimes\Lambda_s$ & 172.8 & 17.2 & 0.37 & 0.7 & 2.85\\
 &$\Psi_N^+\otimes\Lambda_p$ & 162.4 & 6.75 & 0.38 & 0.6 & 2.88\\
 &$\Psi_N^-\otimes\Lambda_s$ & 164.5 & 8.9 & 0.42 & 0.4 & 2.88\\
 &$\Psi_N^-\otimes\Lambda_p$& 154.6 & 7.1 & 0.43 & 0.4 & 2.90\\
\end{tabular}
\end{ruledtabular}
\end{table}

\begin{figure}[th]
  \begin{center}
    \includegraphics[width=0.8\hsize]{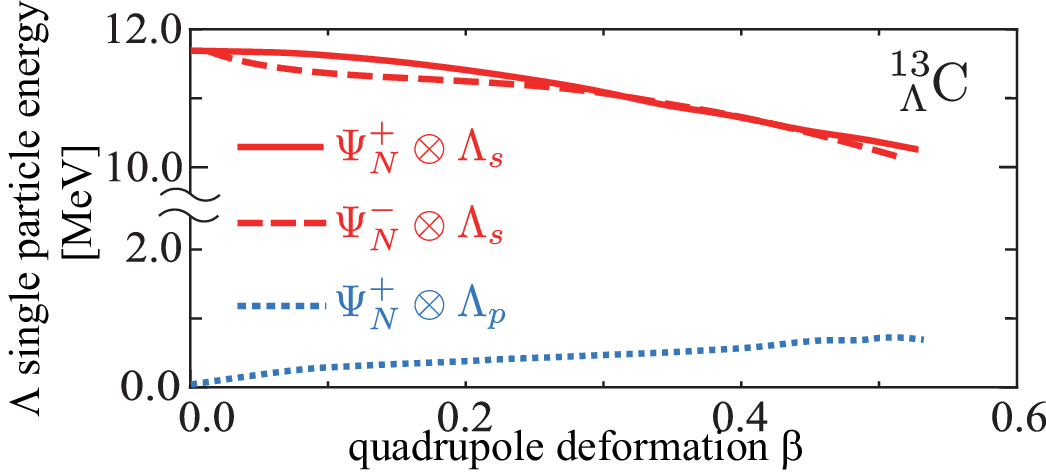}
  \end{center}
  \caption{(color online) The single particle energy of $\Lambda_s$ and $\Lambda_p$ of
 $^{13}_\Lambda$C as function of the quadrupole deformation of core nucleus
 $^{12}$C. Solid (dashed) line shows the  energy of $\Lambda_s$ coupled to the
 positive (negative) parity state of $^{12}$C. Dotted line shows the energy of
 $\Lambda_p$ coupled to the negative-parity state of $^{12}$C.}  
  \label{fig:Lbind}
\end{figure}

\begin{figure}[b]
  \begin{center}
    \includegraphics[width=0.5\hsize]{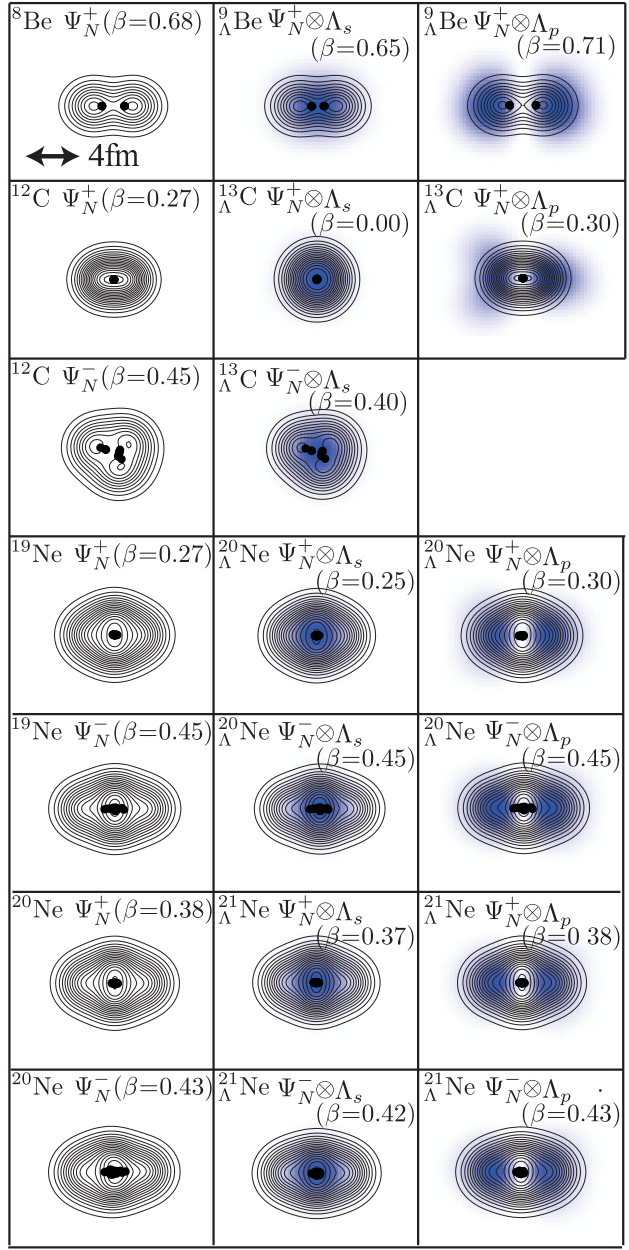}
  \end{center}
  \caption{(color online) Density plot of the intrinsic wave function at each minimum on the
 energy 
 curve. Contour lines shows the density of the nuclear part $\Psi_N$ and the color
 plot shows the $\Lambda$ single particle orbital $\varphi_\Lambda$. Dots in the
 figure shows the centroids of Gaussian wave packets on nuclear part $\Psi_N$.}
  \label{fig:density}
\end{figure}

As the general trend, the shape of energy curve is not strongly modified by
$\Lambda$ particle except for $^{13}_\Lambda$C, and deformation $\beta$ at the
minima are slightly changed. In all cases, $\Lambda_s$  reduces quadrupole
deformation. This is consistent with the cluster model calculations \cite{Yamada84,
Hiyama09} and the (relativistic) mean-field calculations
\cite{Zhou07,Win08,Shulze10} that demonstrated the reduction of $\beta$ by
$\Lambda_s$. On the other hand, it is found that  $\Lambda_p$ increases
$\beta$. The magnitude of the change in quadrupole deformation is strongly dependent
on the core nucleus, and the most prominent in  $^{13}_\Lambda$C in which
$\Lambda_s$ makes $^{12}$C core spherical, while $\Lambda_p$ enhances the core
deformation. The reason of the opposite behavior 
of $\Lambda_s$ and $\Lambda_p$ and the strong dependence on the core nucleus is
clearly seen in the single particle energy of $\Lambda$. Figure \ref{fig:Lbind} shows
the single particle  energy of $\Lambda$ ($\epsilon_\Lambda(\beta)$) in each parity
and   $\Lambda$ single particle state in $^{13}_\Lambda$C. Here
$\epsilon_\Lambda(\beta)$ is defined as the difference between the binding 
energy of $^{13}_\Lambda$C with the deformation $\beta$ and that of corresponding 
state of $^{12}$C with the same deformation, 
\begin{eqnarray}
  \epsilon_\Lambda(\beta) =B_{^{13}_\Lambda\rm C}(\beta)
   - B_{^{12}\rm C}(\beta).
 \end{eqnarray} 
It shows the Nilsson-model-like behavior of the $\Lambda$ single particle
energy. The binding of $\Lambda_s$ becomes shallower as deformation becomes
larger. In the case of $\Lambda_p$, $\Lambda$ occupies the lowest $p$-wave that
comes down as deformation becomes larger. Therefore, $\Lambda_s$ makes quadrupole
deformation smaller and $\Lambda_p$ in the lowest $p$-wave makes it larger. The
$\Lambda$ single particle energy varies within a range of 1$\sim$2 MeV as function
of quadrupole deformation that is smaller than the variation of the core nucleus
energy. It is also case for other calculated hypernuclei. This explains why only
$^{13}_\Lambda$C (FIG. \ref{fig:energy_curve} (b)) manifests the drastic change in 
quadrupole deformation. Since the positive-parity state of $^{12}$C is quite
soft against the quadrupole deformation,  small change in the $\Lambda$ single
particle energy  can result in the large modification in quadrupole deformation. In
other cases, change in $\Lambda$ single particle energy cannot overcome much larger
variation of the core energy and results in minor modification of quadrupole
deformation. Therefore, we can conclude that the drastic change of quadrupole
deformation by $\Lambda$ particle occurs when the core nucleus is quite soft against  
quadrupole deformation within a range of 1$\sim$2 MeV. Since the behavior of the energy
curve is sensitive to the effective NN interaction \cite{Mar06}, the drastic change in
$^{13}_\Lambda$C 
may depend on the choice of NN interaction and it will be investigated in our future work.
The behavior of the $\Lambda$ single particle energy is also understood from the
density distribution of the $\Lambda$ particle and the core nucleus as shown in
Figure \ref{fig:density}. It shows that as nuclear deformation becomes larger the
overlap between the $\Lambda_s$ and the core wave function becomes smaller
(for example, compare $\Psi_N^+\otimes\Lambda_s$ and $\Psi_N^-\otimes\Lambda_s$ of
$^{20}_\Lambda$Ne). It leads to the reduction of $\Lambda$N attraction. On the
contrary, larger deformation makes the overlap larger in the case of $\Lambda_p$
and increases $\Lambda$N attraction (see $\Psi_N^+\otimes\Lambda_p$ and
$\Psi_N^-\otimes\Lambda_p$ of $^{20}_\Lambda$Ne). In the case of $\Lambda_p$,
larger deformation reduces the kinetic energy that also contribute to
the deeper binding of $\Lambda_p$.

Another issue to be mentioned is the reduction of the nuclear radius by $\Lambda$
particle. In all cases, the radius of nuclear part is reduced, but the reduction
(less than 5\%) is much smaller than in the case of $^{7}_\Lambda$Li (20\%)
\cite{Selove88}. More detailed discusstion will be made in our next work.

\subsection{discussion on each hypernucleus}
The calculated total binding energies of $^{8}$Be and $^{9}_\Lambda$Be underestimate
the observed values by about 5MeV. The underestimation is common to all other
hypernuclei. It will be resolved by performing the angular momentum projection and
the  generator coordinate method (GCM) that are usually performed in the study of
normal nuclei by AMD. Indeed, in the case of $^{20}$Ne, it was shown that AMD
calculation \cite{Kimura04} reproduces the observed binding energy. The angular
momentum projection and GCM will be performed in our next work. Despite of the
underestimatin of the total binding energy, $B_\Lambda$ of $\Lambda_p$ is comparable 
with the observed value. 

The density distribution of $\Lambda_p$ in $^{9}_\Lambda$Be
($\Psi_N^+\otimes\Lambda_p$ in FIG. \ref{fig:density}) clearly shows that this state
corresponds to the supersymmetric (genuine) hypernuclear state 
\cite{Bertini81-1, Has98, Hashimoto06}. The nuclear part has the pronouced 2$\alpha$ cluster 
structure and the $\Lambda$ occupies the $p$ orbital parallel to the symmetry
axis. It is also interesting to note that $\Lambda_s$ state reduces the
inter-cluster distance, while $\Lambda_p$ state increases it.

$^{13}_\Lambda$C manifests the drastic change in the quadrupole deformation. The
$\Lambda_s$ makes  $^{12}$C core spherical, while $\Lambda_p$ state enhances
deformation. It is noted that $^{12}$C has the $0p_{3/2}$ subshell closure
configuration at small deformation and 3$\alpha$ cluster structure develops as
deformation becomes larger. In other words, the nucleon spin is not satureted at
small deformation, while that is alomost zero at larger deformation.  The
sophisticated AMD calculation \cite{Enyo98} has shown that the low-lying states of
$^{12}$C have mixed nature between the $0p_{3/2}$ subshell closure configuration and
3$\alpha$ cluster structure, and the mixing strength is different for each state.
Since $\Lambda$ particle changes the deformation and spin property of
$^{12}$C, it will have influence on the $\Lambda$N spin-orbit splitting of
$^{13}_\Lambda$C \cite{Hiyama00,Ajimura01,Millener01}.  

Based on the cluster model calculation, the parity inversion in $^{20}_\Lambda$Ne
was suggested by Sakuda and Bando \cite{Sakuda87}. The core nucleus $^{19}$Ne has the
$\alpha$+$^{15}$O cluster state ($J^\pi$=$1/2^-_1$) 238 keV above the ground state 
($J^\pi$=$1/2^+$) that has $(sd)^3$ shell structure \cite{Sakuda79}. They concluded
that $\Lambda_s$ 
coupled to the $J^\pi$=$1/2^-_1$ state was more deeply bound than that coupled to
the gound state, and the $J^\pi$=$1/2^-\otimes \Lambda_s$ configuration 
becomes the ground state of $^{20}_\Lambda$Ne. They argued that the
$J^\pi$=$1/2^-_1$ state has dilute  $\alpha$+$^{15}$O cluster structure and by the
reduction of the inter-cluster distance, $\Lambda_s$ gains larger binding energy
than the $J^\pi$=$1/2^+_1\otimes\Lambda_s$ configuration. Our result shows the
opposite trend  to their result. Since the positive-parity state is more deformed
than the negative-parity state, the binding of $\Lambda_s$ is weaker when it coupled  
to the negative-parity state. This trend is common to other calculations
including the cluster model calculation for $^{21}_\Lambda$Ne
\cite{Yamada84}. However, AMD fails to 
reproduce small excitation energy of the negative-parity state and it does not have
prominent $\alpha$+$^{15}$O clustering, that are mainly due to the lack of the
angular momentum projection and the GCM calculation. We will need more sophisticated
AMD calculation to settle down this problem. 

The negative-parity state of $^{20}$Ne has larger deformation than the
positive-parity state. Therefore, the $\Lambda_s$ coupled to the positive-parity
state is more deeply bound than that coupled to the negative-parity state. It is
common to other hypernuclei studied here. On the contrary, $\Lambda_p$ is more
deeply bound to the negative-parity state. Since 
number of nucleons in $^{20}$Ne is large enough to  bound $\Lambda_p$, we can expect
that the addition of $\Lambda$  will 
generate a variety of bound rotational bands in $^{21}_\Lambda$Ne as disscussed by
Yamada {\it et al.}\cite{Yamada84}. We will discuss $^{21}_\Lambda$Ne in detail in the
forthcoming paper.

\section{Summary}
An extended version of AMD named HyperAMD has been introduced to investigate
structure of $p$-$sd$ shell hypernuclei. The energy curves of
$^9_\Lambda$Be, $^{13}_\Lambda$C, $^{20}_\Lambda$Ne and $^{21}_\Lambda$Ne as
functions of quadrupole deformation are studied. It has been found that $\Lambda_s$
reduces nuclear deformation, while $\Lambda_p$ increases it. It is due to the
variation of the single particle energy of $\Lambda$ as function of quadrupole
deformation. The binding of $\Lambda_s$ decreases as deformation becomes larger,
while that of $\Lambda_p$ increases. The variation of $\Lambda$ single particle
energy is within a range of 1$\sim$2 MeV, that is rather small compared to the
variation of the energy of the core nucleus. Therefore, the magnitude of the change
of deformation strongly depends on the softness of the core nucleus against
quadrupole deformation. Since $^{12}$C is very soft against quadrupole deformation,
it manfests the most prominent change of quadrupole deformation. This trend of the
deformation change caused by $\Lambda_s$ and $\Lambda_p$ contradicts to the cluster
model calculation for $^{20}_\Lambda$Ne \cite{Sakuda87}, but is consistent with other
calculations. More sphisticated AMD calculation will be needed to resolve this
disagreement.

\bibliography{apssamp}

\end{document}